\documentclass [12pt,a4paper]{article}
\usepackage {amsmath}
\usepackage {graphicx}
\graphicspath{{images/}}
\usepackage {comment}
\usepackage{eqnarray}
\usepackage{relsize}
\usepackage {graphicx}
\graphicspath{{images/}}
\usepackage {comment}
\usepackage{authblk}
\usepackage{cite}
\usepackage{multicol}
\usepackage{longtable}
\usepackage{float}
\restylefloat{table}
\usepackage{multirow}  
\usepackage{amssymb}  
\newcommand \bprime {\backprime\hspace{-.11em}      }   
\date{}
\title{Novel weak form quadrature elements for second strain gradient Euler-Bernoulli beam theory}

\author{Md Ishaquddin\thanks{Corresponding author: \textit{E-mail address: ishaq.isro@gmail.com}}, S.Gopalakrishnan\thanks {\textit{E-mail address: krishnan@iisc.ac.in; Phone: +91-80-22932048}}}

\begin{document}

\maketitle
\vspace{-10mm}
\noindent\textit{Department of Aerospace Engineering, Indian Institute of Science 
Bengaluru 560012, India}\\

\begin{abstract}

Two novel version of weak form quadrature elements are proposed based on Lagrange and Hermite interpolations, respectively, for a second strain gradient Euler-Bernoulli beam theory. The second strain gradient theory is governed by eighth order partial differential equation with displacement, slope, curvature and triple derivative of displacement as degrees of freedom. A simple and efficient differential quadrature frame work is proposed herein to implement these classical and non-classical degrees of freedom. A novel procedure to compute the modified weighting coefficient matrices for the beam element is presented. The proposed elements have displacement as the only degree of freedom in the element domain and displacement, slope, curvature and triple derivative of displacement at the boundaries. The Gauss-Lobatto-Legender quadrature points are assumed as element nodes and also used for numerical integration of the element matrices. Numerical examples are presented to demonstrate the efficiency and accuracy of the proposed beam element.\\

\end{abstract}
\section*{1.0 INTRODUCTION}

The differential quadrature method is an efficient numerical technique for the solutions of linear and non-linear partial differential equations \cite{ Wangb}-\cite{Shu}. It was first introduced by Bellman et.al \cite{Bellman} and later developed by many researchers \cite{Bert1}-\cite{Zhong}. The developments in this field primarily focused on the generalization of this techniques from the perspective of multi-boundary conditions and complicated geometries. As a consequence many improved versions have been developed and applied to the problems of science and engineering \cite{Karami1}-\cite{Wang4}. As exhaustive survey of the developments in the differential quadrature method can be found in the review paper by Malik et.al \cite{review}.  

In this paper we propose for the first time two novel versions of quadrature beam elements based on the weak form of governing equation to solve a eighth order partial differential equation associated with the second strain gradient elasticity theory. The two elements are formulated using the Hermite and Lagrange interpolations, respectively. These elements are the extension of the earlier work by authors \cite{ishaqsf}-\cite{ishaqwf}, on weak form quadrature elements for first strain gradient theory, which is governed by sixth order partial differential equation \cite{Mindlin1}-\cite{Pegios}. The second strain gradient elasticity theory is an enriched version of the classical theory or the first strain gradient theory accounting for higher order gradients of strains, and yiled curvature and triple derivative of displacements as additional degrees of freedom \cite{Mssg}-\cite{Assg}. A novel differential quadrature framework is proposed herein to account for these non-classical degrees of freedom in a simple and efficient manner. This framework is formulated with the aid of variation principles, differential quadrature rule and Gauss Lobatto Legendre (GLL) quadrature rule. Here, the GLL points are used as element nodes and also to perform numerical integration to evaluate the element matrices. The procedure for computing the stiffness matrices at the integration points is analogous to the conventional finite element method. The proposed quadrature elements have displacement, slope, curvature and triple derivative of displacement as the degrees of freedom at the element boundaries and only displacement in the domain. Numerical results on bending, free vibration and stability analysis of second gradient beams are presented to demonstrate the capability of the proposed elements. 

\section{Second strain gradient elasticity theory}

The second strain gradient micro-elasticity theory with two classical and two non-classical material constants is consider in the present study \cite{Mssg,Assg}. The two classical material coefficients are Lam$e^{'}$ constants and the non-classical ones are gradient coefficients with the dimension of length. In what follows, the variational formulation for the second strain gradient Euler-Bernoulli beam theory is presented for the first time. Further, the governing equation and associated classical and non-classical boundary conditions are discussed. \\

The potential energy density function for a second strain gradient theory is given by:

\begin{align*}
W&= \frac{1}{2}\sigma_{ij}\varepsilon_{ij}+\frac{1}{2}g_{1}^{2}(\partial_{k}\sigma_{ij})(\partial_{k}\varepsilon_{ij})+\frac{1}{2}g_{2}^{4}(\partial_{l}\partial_{k}\sigma_{ij})(\partial_{l}\partial_{k}\varepsilon_{ij}) \tag{1}
\end{align*}

The stress-strain relations for 1-D second strain gradient elastic theory are defined as \cite{Mssg}
\begin{align*}
{\tau}&= 2\,\,\mu \,\, \varepsilon + \lambda \,\, {\text{tr}} \varepsilon \,\, \text{I} \\
{{\varsigma}}&= g_{1}^{2} \,\, [2 \,\, \mu \,\, \nabla\varepsilon + \lambda \,\, \nabla (\text{tr}\varepsilon) \,\, \text{I}] 
 \\
{\bar{\varsigma}}&= g_{2}^{4} \,\, [2 \,\, \mu \,\, \nabla\nabla\varepsilon + \lambda \,\, \nabla\nabla (\text{tr}\varepsilon) \,\, \text{I}]  \tag{2}
\end{align*}

\noindent where, $\lambda$,$\,$ $\mu$ are Lam$e^{'}$ constants and $g_{1}$, $g_{2}$ are the strain gradient coefficients of dimension length. $\nabla=\frac{\partial}{\partial x}+\frac{\partial}{\partial y}$ is the Laplacian operator and $\text{I}$ is the unit tensor. $\tau$, $\varsigma$ and  $\bar{\varsigma}$ denotes Cauchy, double and triple stresses respectively, $\varepsilon$ and ($\text{tr}\,\varepsilon$) are the classical strain and its trace which are expressed in terms of displacement vector $\textit{w}$ as:
\begin{align*}
&{\varepsilon}= \frac{1}{2}(\nabla\textit{w}+\textit{w}\nabla)\,\,, \,\,\quad \text{tr}{\varepsilon}= \nabla\textit{w} \tag{3}
\end{align*}

It follows from the above equations the constitutive relations for an Euler-Bernoulli second gradient beam can be stated as
\begin{align*}         
{\tau_{x}}= E\varepsilon_{x}, \quad \varsigma_{x}={g}_{1}^{2}\,\,E\,\varepsilon_{x}^{'}, \quad \bar{\varsigma}_{x}={g}_{2}^{4}\,\,E\,\varepsilon_{x}^{''} \\ \varepsilon_{x}=-z\dfrac{\partial^{2} w(x,t)}{\partial{x}^2}\,\,\,\,\,\,\,\,\,\,\,\,\,\,\,\,\,\,\,\,\,\,\,\,\,\, \tag{4} 
\end{align*}

\noindent For the above state of stress and strain the strain energy expression in terms of displacement for a beam defined over a domain $-L/2\leq x \leq L/2$ can be written as:
\begin{align*}         
{U}= \frac{1}{2}\int_{-L/2}^{L/2} EI\big[(w^{''})^{2}+g_{1}^{2}(w^{'''})^{2}+g_{2}^{4}(w^{\backprime\backprime\prime})^{2}\big]dx-\frac{1}{2}\int_{-L/2}^{L/2} P(w^{'})^{2}dx  \tag{5}
\end{align*}

The potential energy of the applied load is given by
\begin{align*}         
{W}= \int_{-L/2}^{L/2}q(x)w{dx}-\big[Vw\big]_{-L/2}^{L/2}+\big[M{w}^{'}\big]_{-L/2}^{L/2}+\big[\bar{M}{w}^{''}\big]_{-L/2}^{L/2}+\big[\bar{\bar{M}}{w}^{'''}\big]_{-L/2}^{L/2}  \tag{6}
\end{align*}

The kinetic energy is given as
\begin{align*}         
{K}= \frac{1}{2}\int_{t_{0}}^{t_{1}}\int_{-L/2}^{L/2}\rho{A}\dot{w}^{2}{dx}{dt}  \tag{7}
\end{align*}

\noindent  where, $E$, $A$ and $I$ are the Young's modulus, area, moment of inertia, respectively. $q$ and $w(x,t)$ are the transverse load and displacement of the beam. $V$, $M$, $\bar{M}$ and $\bar{\bar{M}}$ are the shear force, bending moment, double and triple moment, respectively. $P$ is the axial compressive force and over dot indicates differentiation with respect to time.\\
 
Using the The Hamilton's principle\cite{Reddyb}:
\begin{align*}         
\delta\int_{t_{0}}^{t_{1}}(U-W-K)\,dt=0   \tag{8}
\end{align*}

\noindent we get the following weak form expression for elastic stiffness matrix `K', geometric stiffness matrix 'G' and consistent mass matrix `m' as 
\begin{align*}   \label{K_beam}      
&K= \int_{-L/2}^{L/2} EI\big[w{''}\,\delta{w{''}}+g_{1}^{2}	\,w^{'''}\delta{w}^{'''}+g_{2}^{4}	\,w^{\backprime\backprime\prime}\delta{w}^{\backprime\backprime\prime}\big]dx \tag{9}
\end{align*}
\begin{align*}   \label{G_beam}      
&G= \int_{-L/2}^{L/2} Pw{'}\,\delta{w{'}}dx \tag{10}
\end{align*}
\begin{align*}   \label{M_beam}      
 &m=\int_{-L/2}^{L/2}\,\rho{A}\,\dot{w}\,\dot{\delta{w}}\,dx \tag{11}
\end{align*}

\noindent Using the Equations (\ref{K_beam})-(\ref{M_beam}) the governing equation of motion for a second strain gradient Euler-Bernoulli beam is obtained as

\begin{align*}   \label{EOM_beam}      
EI(w^{\backprime\backprime\prime}-g_{1}^{2}w^{\backprime\prime\prime}+
g_{2}^{4}w^{\backprime\prime\prime\prime\prime})+q+Pw^{''}+\rho{A}\ddot{w}=0 \tag{12}
\end{align*}

\noindent and the associated boundary conditions are: \\

\noindent \textit{Classical} :     
\begin{align*}  \label{eq:BC_Cl_Beam}     
V&=EI[w^{'''}-g_{1}^{2}w^{\bprime\prime}+g_{2}^{4}w^{\bprime\prime\prime\prime}]=0 \hspace{0.5cm}\text{or}\hspace{0.5cm} w=0,\hspace{0.5cm}\text{at}\,\,x=(-L/2,L/2)\\
 M&=EI[w^{''}-g_{1}^{2}w^{\prime\bprime\prime}+g_{2}^{4}w^{\bprime\prime\prime}]=0 \hspace{0.5cm}\text{or} \hspace{0.5cm}w^{'}=0,\hspace{0.5cm}\text{at}\,\,x=(-L/2,L/2)
\tag{13}
\end{align*}

\noindent \textit { Non-classical} :         
\begin{align*}  \label{eq:BC_NCl_Beam}       
\bar{M}&=EI[g_{1}^{2}w^{'''}-g_{2}^{4}w^{\backprime\prime\prime}]=0 \hspace{0.5cm}\text{or}\hspace{0.5cm} w^{''}=0,\,\,\,\text{at}\,\,x=(-L/2,L/2)		\\
\bar{\bar{M}}&=EI\,{g}_{2}^{4}\,w^{\backprime\backprime\prime}=0 \hspace{2.2cm}\text{or}\hspace{0.5cm} w^{'''}=0,\,\,\,\text{at}\,\,x=(-L/2,L/2)		\tag{14}
\end{align*} 
\section{Quadrature elements for a second strain gradient Euler-Bernoulli beam} \label{beam_qem}

\begin{figure}[H]
\includegraphics[width=1.0\textwidth]{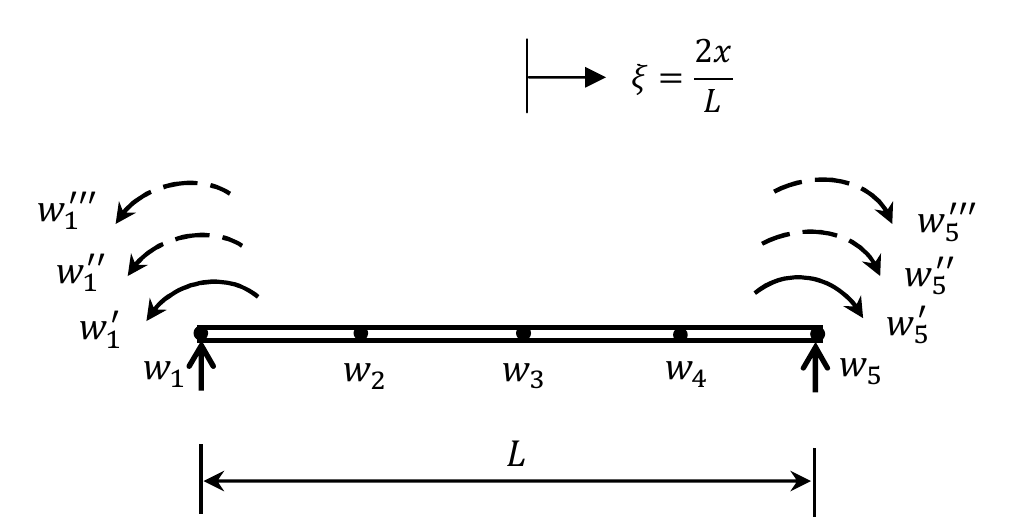}
\centering
\caption{A typical quadrature element for a second strain gradient Euler-Bernoulli beam.}
\label{fig:beam}
\end{figure}

In this section we present, two novel quadrature elements for a second gradient Euler-Bernoulli beam. The first element is based on Lagrangian interpolation and the later is formulated using Hermite interpolation. A typical N-node quadrature element for an Euler-Bernoulli second strain gradient beam is shown in the Figure \ref{fig:beam}. 
It can be observed that each interior node has only displacement $w$ as degrees of freedom and the boundary has 4 degrees of freedom $w$, $w^{'}$, $w^{''}$ and $w^{'''}$. The displacement vector includes the slope, curvature and triple displacement derivative as additional degrees of freedom at the element boundaries given by: $w=\{w_{1},\cdots, w_{N},w^{'}_{1},w^{'}_{N},w^{''}_{1},w^{''}_{N},w^{'''}_{1},w^{'''}_{N}\}$. The procedure to incorporate these extra boundary degrees of freedom while formulating the Lagrange and Hermite interpolation based quadrature elements will be present next.

\subsection{Lagrange interpolation based quadrature beam element} \label{section_Lag_qem_beam}
The displacement for a N-node quadrature beam is assumed as\cite{Wangb}:
\begin{align*}    \label{disp_Lag_qem_beam}
w(x,t)=\sum_{j=1}^{N} L_{j}(x)w^{b}_{j}=\sum_{j=1}^{N} \bar{L}_{j}(\xi)w^{b}_{j} \tag{15}
\end{align*}

\noindent $L_{j}(x)$ and $\bar{L}_{j}(\xi)$ are the conventional Lagrangian interpolation functions in $x$ and $\xi$ co-ordinates respectively, and $\xi=2x/L$ with $\xi \in[-1,1]$. The Lagrange interpolation functions can be defined as\cite{Wangb,Shu}
\begin{align*}    
L_{j}(\xi)=\frac{\beta(\xi)}{\beta(\xi_{j})}=\prod_{\substack{k=1 \\ (k\neq j)}}^{N}\frac{(\xi-\xi_{k})}{(\xi_{j}-\xi_{k})}   \tag{16}
\end{align*}
\noindent where \\
 $\beta(\xi)=(\xi-\xi_{1})(\xi-\xi_{2})\cdots(\xi-\xi_{j-1})(\xi-\xi_{j+1})\cdots(\xi-\xi_{N})$ \\
 $\beta(\xi_{j})=(\xi_{j}-\xi_{1})(\xi_{j}-\xi_{2})\cdots(\xi_{j}-\xi_{j-1})(\xi_{j}-\xi_{j+1})\cdots)(\xi_{j}-\xi_{N})$ \\

\noindent The first order derivative of the Lagrange interpolation function is obtained as,
\begin{align*}    
A_{ij}={L}^{'}_{j}(\xi_{i})\begin{cases}
\mathlarger\prod_{\substack{k=1 \\ (k\neq i,j)}}^{N}(\xi_{i}-\xi_{k})/\mathlarger\prod_{\substack{k=1 \\ (k\neq j)}}^{N}=(\xi_{j}-\xi_{k})\,\,\,\, (i\neq j)\\ \\ 
\mathlarger\sum_{\substack{k=1 \\ (k\neq i)}}^{N}\frac{1}{(\xi_{i}-\xi_{k})}
\end{cases}\tag{17}  
\end{align*}

The higher order derivatives of Lagrange interpolation functions are obtained as
\begin{align*}    
B_{ij}=\sum_{k=1}^{N} A_{ik}A_{kj} \,\, , \quad
C_{ij}=\sum_{k=1}^{N} B_{ik}A_{kj}  \,\,\,(i,j=1,2,...,N) \\ 
D_{ij}=\sum_{k=1}^{N} B_{ik}B_{kj}  \,\,\,(i,j=1,2,...,N)\,\,\,\,\,\,\,\,\,\,\,\,\,\,\,\,\,\,\,\,\tag{18}
\end{align*}\\
Here, $B_{ij}$, $C_{ij}$ and $D_{ij}$ are weighting coefficients of Lagrange interpolation functions for second, third and fourth order derivatives, respectively.

The eighth order partial differential equation given in Equation (\ref{EOM_beam}), renders  slope $w^{'}$, curvature $w^{''}$ and triple displacement derivative $w^{'''}$ as extra degrees of freedom at the element boundaries. To account for these extra boundary degrees of freedom in the formulation, the derivatives of conventional weighting function $A_{ij}$, $B_{ij}$, $C_{ij}$ and $D_{ij}$ are modified as follows:\\ 

\noindent \textit{First order derivative matrix}: 
\begin{align*}    \label{Lag_Aij}
\bar{A}_{ij}=\begin{cases}
A_{ij} \,\,\,\, (i,j=1,2,\cdots,N)\\ \\
0 \,\,\,\,\,\,\,(i,j=1,2,\cdots,N;\,\,\,\,j=N+1,\cdots,N+6)\tag{19}
\end{cases} 
\end{align*}
\noindent \textit{Second order derivative matrix}: 
\begin{align*} \label{Lag_Bij1}
\bar{B}_{ij}=\begin{cases}
B_{ij} \,\,\,\, (j=1,2,\cdots,N)\\ \\
0 \,\,\,\,\,\,\,(j=N+1,\cdots,N+6;\,\,\,\, i=2,3,\cdots,N-1)\tag{20}
\end{cases} 
\end{align*}
\begin{align*} \label{Lag_Bij2}
\bar{B}_{ij}=\sum_{k=2}^{N-1}A_{ik}A_{kj} \,\,\, (j=1,2,\cdots,N;\,\,i=1,N)\\ 
\bar{B}_{i(N+1)}=A_{i1}\,\,; \,\, \,\,\,  \,\,\bar{B}_{i(N+2)}=A_{iN} \,\,\,\,(i=1,N)\tag{21}
\end{align*} 

\noindent \textit{Third order derivative matrix}: 
\begin{align*}\label{Lag_Cij1}
\bar{C}_{ij}=\begin{cases}
C_{ij} \,\,\,\, (j=1,2,\cdots,N)\\ \\
0 \,\,\,\,\,\,\,(j=N+1,\cdots,N+6;\,\,\,\, i=2,3,\cdots,N-1)\tag{22}
\end{cases}
\end{align*} 
\begin{align*} \label{Lag_Cij2}
\bar{C}_{ij}=\sum_{k=2}^{N-1}B_{ik}A_{kj}\,\,\  (j=1,2,\cdots,N;\,\,\,\,i=1,N)\\ 
\bar{C}_{i(N+3)}=A_{i1}\,\,; \,\, \,\, \,\,\, \bar{C}_{i(N+4)}=A_{iN} \,\,\,\,(i=1,N)\tag{23}
\end{align*} 

\noindent \textit{Fourth order derivative matrix}: 
\begin{align*}\label{Lag_Dij1}
\bar{D}_{ij}=\begin{cases}
D_{ij} \,\,\,\, (j=1,2,\cdots,N)\\ \\
0 \,\,\,\,\,\,\,(j=N+1,\cdots,N+6;\,\,\,\, i=2,3,\cdots,N-1)\tag{24}
\end{cases}	
\end{align*} 
\begin{align*} \label{Lag_Dij2}
\bar{D}_{ij}=\sum_{k=2}^{N-1}B_{ik}B_{kj}\,\,\  (j=1,2,\cdots,N;\,\,\,\,i=1,N)\\ 
\bar{D}_{i(N+5)}=A_{i1}\,\,; \,\, \,\, \,\,\, \bar{D}_{i(N+6)}=A_{iN} \,\,\,\,(i=1,N)\tag{25}
\end{align*} 

 Using the above Equations (\ref{Lag_Aij})-(\ref{Lag_Dij2}), the element matrices can be expressed in terms of weighting coefficients as \\
	 
\noindent $Elastic \,stiffness\, matrix$ :
\begin{align*}     \label{Lag_Kij_dqm_beam}    
K_{ij}=\frac{8EI}{L^{3}}\sum_{k=1}^{N} {H}_{k}\bar{B}_{ki}\bar{B}_{kj}+g_{1}^{2}\frac{32EI}{L^{5}}\sum_{k=1}^{N} {H}_{k}\bar{C}_{ki}\bar{C}_{kj}+g_{2}^{4}\frac{128EI}{L^{7}}\sum_{k=1}^{N} {H}_{k}\bar{D}_{ki}\bar{D}_{kj} \\ \,\,\,\,\,
(i,j=1,2,...,N,N+1,\cdots,N+6)\tag{26}
\end{align*}

\noindent $Geometric \,stiffness\, matrix$ :
\begin{align*}    \label{Herm_Gij_dqm_beam}
G_{ij}=P\sum_{k=1}^{N} {H}_{k}A_{ki}A_{kj} \,\,\,\,\,\,\,\,\,\,\,\,\,\,\,\,\,\,\,\,\,\,\,\,\,\,\,\,\,\,\\ 
\hspace{0.6cm}(i,j=1,2,...,N,N+1,\cdots,N+6)\tag{27}
\end{align*}
\noindent $Consistent\, mass\, matrix$ :
\begin{align*}     \label{Lag_Mij_dqm_beam}     
M_{ij}=\frac{\rho{A}L}{2}{H}_{i}\delta_{ij}\,\,\,
(i,j=1,2,...,N)\tag{28}
\end{align*}

\noindent \textit{Equivalent load vector}: 
\begin{align*}   \label{load_beam}      
&f_{i}= \frac{L}{2}\,q(\xi)\,H_{i} \,\,\,\ (i=1,2,...,N) \tag{29}
\end{align*}

Here $\xi$ and $H$ are the coordinate and weights of GLL quadrature.      
 $\delta_{ij}$ is the Dirac-delta function. 

\subsection{Hermite interpolation based quadrature beam element} \label{section_Herm_qem_beam}

The displacement for a N-node second strain gradient beam element based on Hermite interpolations is assumed as
\begin{align*}    \label{disp_Herm_qem_beam}
w(\xi,t)=\sum_{j=1}^{N} \phi_{j}(\xi)w_{j}+\psi_{1}(\xi)w_{1}^{'}+\psi_{N}(\xi)w_{N}^{'}+\varphi_{1}(\xi)w_{1}^{''}+\varphi_{N}(\xi)w_{N}^{''}+\\
\zeta_{1}(\xi)w_{1}^{'''}+\zeta_{N}(\xi)w_{N}^{'''}=\sum_{j=1}^{N+6} \Gamma_{j}(\xi)w_{j}\,\,\,\,\,\,\,\,\,\,\,\,\,\,\,\,\,\,\,\,\,\,\,\,\,\,\,\,\,\,\,\,\,\,\,\,\,\,\,\,\,\,\,\,\,\,\,\,\,\,\,\,\,\,\,\,\,\,\,\,\,\,\,\tag{30}
\end{align*}

\noindent $\phi$, $\psi$, $\varphi$ and $\zeta$ are Hermite interpolation functions defined as \cite{Eigth}

\begin{align*}    \label{Herm_zeta}
\zeta_{j}(\xi)=\frac{1}{6(\xi_{j}-\xi_{N-j+1})^{3}}L_{j}(\xi)(\xi-\xi_{j})^{3}(\xi-\xi_{N-j+1})^{3}  \,\,\,\,\,\,\, (j=1, N)\tag{31}
\end{align*}

\begin{align*}    \label{Herm_varphi}
\varphi_{j}(\xi)=\frac{1}{(\xi_{j}-\xi_{N-j+1})^{3}}L^{1}_{j}(\xi)(\xi-\xi_{N-j+1})^{3}(a_{j3}\,\xi^{3}+b_{j3}\,\xi^{2}+c_{j3}\,\xi+d_{j3})  \,\,\, (j=1, N)\tag{32}
\end{align*}

\begin{align*} 
\noindent \text{where},\,\,  a_{j3}=-\frac{1.5}{(\xi_{j}-\xi_{N-j+1})}-\frac{L^{1}_{j}(\xi_{j})}{2},   \quad b_{j3}=\frac{1}{2}-3a_{j3}\xi_{j},\,\,\,\,\,\,\,\,\,\,\,\,\,\,\,\,\,\,\,\,\,\,\,\, \quad \\ c_{j3}=-3a_{j3}\xi_{j}^{2}-2b_{j3}\xi_{j}, \quad  d_{j3}=-a_{j3}\xi_{j}^{3}-b_{j3}\xi_{j}^{2}-c_{j3}\xi_{j} \,\,\,\,\,\,\,\,\,\,\,\,\,\,\,\,\,\,\,\,\,\,\,\,
\end{align*}

\begin{align*}    \label{Herm_sie}
\psi_{j}(\xi)=\frac{1}{(\xi_{j}-\xi_{N-j+1})^{3}}L^{1}_{j}(\xi)(\xi-\xi_{N-j+1})^{3}(a_{j2}\,\xi^{3}+b_{j2}\,\xi^{2}+c_{j2}\,\xi+d_{j2})  \,\,\, (j=1, N)\tag{33}
\end{align*}
\begin{align*}    
\text{where},\,\, a_{j2}= \frac{6}{(\xi_{j}-\xi_{N-j+1})^{2}}+\frac{3L^{1}_{j}(\xi_{j})}{(\xi_{j}-\xi_{N-j+1})}-\frac{L^{2}_{j}(\xi_{j})}{2}+[L^{1}_{j}(\xi_{j})]^{2} \,\,\,\,\,\,\,\,\,\,\,\,\,\,\,\,\,\,\,\,\,\, \\
b_{j2}=-\frac{3}{(\xi_{j}-\xi_{N-j+1})}-3a_{j2}\xi_{j}-L^{1}_{j}(\xi_{j}),\,\,  c_{j2}=1-3a_{j2}\xi_{j}^{2}-2b_{j2}\xi_{j},\\ d_{j2}=-a_{j2}\xi_{j}^{3}-b_{j2}\xi_{j}^{2}-c_{j2}\xi_{j}\,\,\,\,\,\,\,\,\,\,\,\,\,\,\,\,\,\,\,\,\,\,\,\,\,\,\,\,\,\,\,\,\,\,\,\,\,\,\,\,\,\,\,\,\,\,\,\,\,\,\,\,\,\,\,\,\,\,\,\,\,\,\,\,\,\,\,\,\,\,\,\,\,\,\,\,\,\,\,\,\,\,\,\,\,\,\,\,\,\,\,\,\,\,\,\,\,    
\end{align*}
\begin{align*}    \label{Herm_phi}
\phi_{j}(\xi)=\frac{1}{(\xi_{j}-\xi_{N-j+1})^{3}}L^{1}_{j}(\xi)(\xi-\xi_{N-j+1})^{3}(a_{j1}\,\xi^{3}+b_{j1}\,\xi^{2}+c_{j1}\,\xi+d_{j1})  \,\,\, (j=1, N)\tag{34}
\end{align*}
\begin{align*}   
\text{where},\,\,\,a_{j1}= \frac{1.5}{(\xi_{j}-\xi_{N-j+1})}\{L^{2}_{j}(\xi_{j})-
2[L^{1}_{j}(\xi_{j})]^{2}\} -\frac{10}{(\xi_{j}-\xi_{N-j+1})^{3}}-\,\,\,\,\,\,\,\,\,\,\,\,\,\,\,\,\\
\frac{6L^{1}_{j}(\xi_{j})}{(\xi_{j}-\xi_{N-j+1})^{2}} -\frac{L^{3}_{j}(\xi_{j})}{6}+L^{1}_{j}(\xi_{j})L^{2}_{j}(\xi_{j})-[L^{1}_{j}(\xi_{j})]^{3}\,\,\,\,\,\,\,\,\,\,\,\,\,\,\,\,\,\,\,\,\,\,\,\,\\
\,\,\,\,\,\,\,\,\,\, b_{j1}=\frac{3L^{1}_{j}(\xi_{j})}{(\xi_{j}-\xi_{N-j+1})}+\frac{6}{(\xi_{j}-\xi_{N-j+1})^{2}}-\frac{L^{2}_{j}(\xi_{j})}{2}
+[L^{1}_{j}(\xi_{j})]^{2}-3a_{j1}\xi_{j} \\
c_{j1}=-\frac{3}{(\xi_{j}-\xi_{N-j+1})}
-L^{1}_{j}(\xi_{j})-3a_{j1}\xi_{j}^{2}-2b_{j1}\xi_{j},\,\,\,\,\,\,\,\,\,\,\,\,\,\,\,\,\,\,\,\,\,\,\,\,\,\,\,\,\,\,\,\,\,\,\,\,\,\,\,\,\,\,\,\,\,\\
d_{j1}=1-a_{j1}\xi_{j}^{3}-b_{j1}\xi_{j}^{2}-c_{j1}\xi_{j}  \,\,\,\,\,\,\,\,\,\,\,\,\,\,\,\,\,\,\,\,\,\,\,\,\,\,\,\,\,\,\,\,\,\,\,\,\,\,\,\,\,\,\,\,\,\,\,\,\,\,\,\,\,\,\,\,\,\,\,\,\,\,\,\,\,\,\,\,\,\,\,\,\,\,\,\,\,\,\,\,\,\,\,\,\,\,\,\,\,\,\,
 \end{align*}
\begin{align*}  \label{Herm_phiD}
\phi_{j}(\xi)=\frac{1}{(\xi_{j}-\xi_{1})^{3}(\xi_{j}-\xi_{N})^{3}}L_{j}(\xi)(\xi-\xi_{1})^{3}(\xi-\xi_{N})^{3}   \,\,\, (j=2,3,...,N-1)\tag{35}
\end{align*}

The $k$th order derivative of $w(\xi)$ with respect to $\xi$ is obtained from Equation (\ref{disp_Herm_qem_beam}) as 
\begin{align*}   \label{Herm_Aij_main} 
w^{k}(\xi)=\sum_{j=1}^{N} \phi_{j}^{k}(\xi)w_{j}+\psi_{1}^{k}(\xi)w_{1}^{'}+\psi_{N}^{k}(\xi)w_{N}^{'}+\varphi_{1}^{k}(\xi)w_{1}^{''}+\varphi_{N}^{k}(\xi)w_{N}^{''}+\\
\zeta_{1}^{k}(\xi)w_{1}^{'''}+\zeta_{N}^{k}(\xi)w_{N}^{'''}=\sum_{j=1}^{N+6} \Gamma_{j}^{k}(\xi)w_{j}\tag{36}
\end{align*}

The derivatives of the Hermite interpolations are defined as follows:
\begin{align*}    \label{Herm_zeta_Der}
\zeta^{k}_{j}(\xi)=\frac{1}{6(\xi_{j}-\xi_{N-j+1})^{3}}\bigg\{L^{k}_{j}(\xi)(\xi-\xi_{j})^{3}(\xi-\xi_{N-j+1})^{3}+ 
3kL^{k-1}_{j}(\xi)\\(\xi-\xi_{j})^{3}(\xi-\xi_{N-j+1})^{2}+ 
3kL^{k-1}_{j}(\xi)(\xi-\xi_{j})^{2}(\xi-\xi_{N-j+1})^{3}+ \\
9k(k-1)L^{k-2}_{j}(\xi)(\xi-\xi_{j})^{2}(\xi-\xi_{N-j+1})^{2}+ 
3k(k-1)L^{k-2}_{j}(\xi)\\(\xi-\xi_{j})^{3}(\xi-\xi_{N-j+1})+ 
3k(k-1)L^{k-2}_{j}(\xi)(\xi-\xi_{j})\\(\xi-\xi_{N-j+1})^{3}+ 
k(k-1)(k-2)L^{k-2}_{j}(\xi)(\xi-\xi_{j})^{3}+\\
k(k-1)(k-2)L^{k-2}_{j}(\xi)(\xi-\xi_{N-j+1})^{3}+\\
9k(k-1)(k-2)L^{k-3}_{j}(\xi)(\xi-\xi_{N-j+1})(\xi-\xi_{j})^{2}+ \\
9k(k-1)(k-2)L^{k-3}_{j}(\xi)(\xi-\xi_{N-j+1})^{2}(\xi-\xi_{j})+\\3k(k-1)(k-2)(k-3)L^{k-2}_{j}(\xi)(\xi-\xi_{j})(\xi-\xi_{N-j+1})^{2}+\\3k(k-1)(k-2)(k-3)L^{k-2}_{j}(\xi)(\xi-\xi_{N-j+1})(\xi-\xi_{j})^{2}+\\9k(k-1)(k-2)L^{k-3}_{j}(\xi)(\xi-\xi_{N-j+1})(\xi-\xi_{j})\bigg\} \\ \,\,\,\,\,\,\,\,\,\,\,\,\,\,\,\, (j=1, N)\tag{37}
\end{align*}
\begin{align*}    \label{Herm_phi_Der}
\phi_{j}^{k}(\xi)=\frac{1}{(\xi_{j}-\xi_{j})^{3}(\xi_{j}-\xi_{N-j+1})^{3}}\bigg\{L^{k}_{j}(\xi)(\xi-\xi_{j})^{3}(\xi-\xi_{N-j+1})^{3}+ 
3kL^{k-1}_{j}(\xi)\\(\xi-\xi_{j})^{3}(\xi-\xi_{N-j+1})^{2}+ 
3kL^{k-1}_{j}(\xi)(\xi-\xi_{j})^{2}(\xi-\xi_{N-j+1})^{3}+ \\
9k(k-1)L^{k-2}_{j}(\xi)(\xi-\xi_{j})^{2}(\xi-\xi_{N-j+1})^{2}+ 
3k(k-1)L^{k-2}_{j}(\xi)\\(\xi-\xi_{j})^{3}(\xi-\xi_{N-j+1})+ 
3k(k-1)L^{k-2}_{j}(\xi)(\xi-\xi_{j})(\xi-\xi_{N-j+1})^{3}+ \\
k(k-1)(k-2)L^{k-2}_{j}(\xi)(\xi-\xi_{j})^{3}+
k(k-1)(k-2)L^{k-2}_{j}(\xi)\\(\xi-\xi_{N-j+1})^{3}+
9k(k-1)(k-2)L^{k-3}_{j}(\xi)(\xi-\xi_{N-j+1})(\xi-\xi_{j})^{2}+ \\
9k(k-1)(k-2)L^{k-3}_{j}(\xi)(\xi-\xi_{N-j+1})^{2}(\xi-\xi_{j})+\\3k(k-1)(k-2)(k-3)L^{k-2}_{j}(\xi)(\xi-\xi_{j})(\xi-\xi_{N-j+1})^{2}+\\3k(k-1)(k-2)(k-3)L^{k-2}_{j}(\xi)(\xi-\xi_{N-j+1})(\xi-\xi_{j})^{2}+\\9k(k-1)(k-2)L^{k-3}_{j}(\xi)(\xi-\xi_{N-j+1})(\xi-\xi_{j})\bigg\} \\ \,\,\,\,\,\,\,\,\,\,\,\,\,\,\,\, (j=2,3,...,N-1)\tag{38}
\end{align*}
\begin{align*}    \label{Herm_Phi_Der_B}
\phi_{j}^{k}(\xi)=\frac{1}{(\xi_{j}-\xi_{N-j+1})^{3}}\bigg\{L^{k}_{j}(\xi)(\xi-\xi_{j})^{3}(a_{j1}\xi^{3}+b_{j1}\xi^{2}+c_{j1}\xi+d_{j1})+ \\
3kL^{k-1}_{j}(\xi)(a_{j1}\xi^{3}+b_{j1}\xi^{2}+c_{j1}\xi+d_{j1})(\xi-\xi_{j})^{2}+\\
kL^{k-1}_{j}(\xi)(3a_{j1}\xi^{2}+2b_{j1}\xi+c_{j1})(\xi-\xi_{j})^{3}+3k(k-1)L^{k-2}_{j}(\xi)\\(3a_{j1}\xi^{2}+2b_{j1}\xi+c_{j1})(\xi-\xi_{j})^{2}+3k(k-1)L^{k-2}_{j}(\xi)(a_{j1}\xi^{3}\\ 
+ b_{j1}\xi^{2}+c_{j1}\xi+d_{j1})(\xi-\xi_{j})+k(k-1)L^{k-2}_{j}(\xi)(3a_{j1}\xi+b_{j1}\xi)\\(\xi-\xi_{j})^{3}+
k(k-1)(k-2)L^{k-3}_{j}(\xi)(a_{j1}\xi^{3}+b_{j1}\xi^{2}+c_{j1}\xi+d_{j1})+\\
3k(k-1)(k-2)L^{k-3}_{j}(\xi-\xi_{j})(3a_{j1}\xi^{2}+2b_{j1}\xi+c_{j1})+\\
3k(k-1)L^{k-3}_{j}(\xi-\xi_{j})^{2}(6a_{j1}\xi+2b_{j1})+k(k-1)\\(k-2)a_{j1}L^{k-3}_{j}(\xi-\xi_{j})^{3}+
k(k-1)(k-2)L_{j}(3a_{j1}\xi^{2}+\\2b_{j1}\xi+c_{j1})+
3k(k-1)(k-2)L_{j}(\xi-\xi_{j})^{2}+\\3k(k-1)L_{j}(\xi-\xi_{j})(6a_{j1}\xi+2b_{j1})\bigg\} \\ \,\,\,\,\,\,\,\,\,\,\,\,\,\,\,\, (j=1,N)\tag{39}
\end{align*}
\begin{align*}    \label{Herm_Gamma}
\gamma_{j}^{k}(\xi)=\frac{1}{(\xi_{j}-\xi_{N-j+1})^{3}}\bigg\{L^{k}_{j}(\xi)(\xi-\xi_{j})^{3}(\alpha_{j1}\xi^{3}+\alpha_{j2}\xi^{2}+\alpha_{j3}\xi+\alpha_{j4})+ \\
3kL^{k-1}_{j}(\xi)(\alpha_{j1}\xi^{3}+\alpha_{j2}\xi^{2}+\alpha_{j3}\xi+\alpha_{j4})(\xi-\xi_{j})^{2}+\\
kL^{k-1}_{j}(\xi)(3\alpha_{j1}\xi^{2}+2\alpha_{j2}\xi+\alpha_{j3})(\xi-\xi_{j})^{3}+3k(k-1)L^{k-2}_{j}(\xi)\\(3\alpha_{j1}\xi^{2}+2\alpha_{j2}\xi+\alpha_{j3})(\xi-\xi_{j})^{2}+3k(k-1)L^{k-2}_{j}(\xi)(\alpha_{j1}\xi^{3}\\ 
+ \alpha_{j2}\xi^{2}+\alpha_{j3}\xi+\alpha_{j4})(\xi-\xi_{j})+k(k-1)L^{k-2}_{j}(\xi)(3\alpha_{j1}\xi+\alpha_{j2}\xi)\\(\xi-\xi_{j})^{3}+
k(k-1)(k-2)L^{k-3}_{j}(\xi)(\alpha_{j1}\xi^{3}+\alpha_{j2}\xi^{2}+\alpha_{j3}\xi+\alpha_{j4})+\\
3k(k-1)(k-2)L^{k-3}_{j}(\xi-\xi_{j})(3\alpha_{j1}\xi^{2}+2\alpha_{j2}\xi+\alpha_{j3})+\\
3k(k-1)L^{k-3}_{j}(\xi-\xi_{j})^{2}(6\alpha_{j1}\xi+2\alpha_{j2})+k(k-1)\\(k-2)\alpha_{j1}L^{k-3}_{j}(\xi-\xi_{j})^{3}+
k(k-1)(k-2)L_{j}(3\alpha_{j1}\xi^{2}+\\2\alpha_{j2}\xi+\alpha_{j3})+
3k(k-1)(k-2)L_{j}(\xi-\xi_{j})^{2}+\\3k(k-1)L_{j}(\xi-\xi_{j})(6\alpha_{j1}\xi+2\alpha_{j2})\bigg\} \\ \,\,\,\,\,\,\,\,\,\,\,\,\,\,\,\, (j=1,N)\tag{40}
\end{align*}

\noindent In the above equations, $\phi_{j}^{k}(\xi)=\gamma_{j}^{k}(\xi)$ with $\alpha_{j1}=a_{j1}, \alpha_{j2}=b_{j1}, \alpha_{j3}=c_{j1}, \alpha_{j4}=d_{j1}$. Similarly, $\psi_{j}^{k}(\xi)=\gamma_{j}^{k}(\xi)$ with $\alpha_{j1}=a_{j2}, \alpha_{j2}=b_{j2}, \alpha_{j3}=c_{j2}, \alpha_{j4}=d_{j2}$ and $\varphi _{j}^{k}(\xi)=\gamma_{j}^{k}(\xi)$ with $\alpha_{j1}=a_{j3}, \alpha_{j2}=b_{j3}, \alpha_{j3}=c_{j3}, \alpha_{j4}=d_{j3}$.

 Using the above Equation (\ref{disp_Herm_qem_beam})-(\ref{Herm_Gamma}), the element matrices can be expressed in terms of weighting coefficients as \\

\noindent $Elastic \,stiffness\, matrix$ :
\begin{align*}    \label{Herm_Kij_dqm_beam}
K_{ij}=\frac{8EI}{L^{3}}\sum_{k=1}^{N} {H}_{k}\Gamma_{ki}^{(2)}\Gamma_{kj}^{(2)}+g_{1}^{2}\frac{32EI}{L^{5}}\sum_{k=1}^{N} {H}_{k}\Gamma_{ki}^{(3)}\Gamma_{kj}^{(3)} +g_{2}^{4}\frac{128EI}{L^{7}}\sum_{k=1}^{N} {H}_{k}\Gamma_{ki}^{(4)}\Gamma_{kj}^{(4)}\\ \,\,\,\,\,
(i,j=1,2,...,N,N+1,\cdots,N+6)\tag{41}
\end{align*}

\noindent $Geometric \,stiffness\, matrix$ :
\begin{align*}    \label{Herm_Gij_dqm_beam}
G_{ij}=P\sum_{k=1}^{N} {H}_{k}\Gamma_{ki}^{(1)}\Gamma_{kj}^{(1)}\,\,\,\,\,\,\,\,\,\,\,\,\,\,\,\, (i,j=1,2,...,N,N+1,\cdots,N+6)\tag{42}
\end{align*}

Here $\xi$ and $H$ are the coordinate and weights of GLL quadrature.      
 $\delta_{ij}$ is the Dirac-delta function. The super-script of $\Gamma$ in the Equation (\ref{Herm_Kij_dqm_beam})-(\ref{Herm_Gij_dqm_beam}) indicates the order of derivative of the Hermite interpolation function.The consistent mass matrix and equivalent load vector remains the same as given by Equation (\ref{Lag_Mij_dqm_beam})-(\ref{load_beam}).

Combining the elastic stiffness, geometric stiffness, mass matrix and equivalent load vector we get the following system of equations

\begin{align*}\label{eq:Boundary_disp_Beam}
\begin{bmatrix}
k_{bb} & \phantom{-}k_{bd} \\ \\ 
k_{db} & \phantom{-}k_{dd} \\ \\ 
 \end{bmatrix}\begin{Bmatrix}
\Delta_{b} \\ \\ 
\Delta_{d}  \\ \\ 
 \end{Bmatrix}=\begin{Bmatrix}
f_{b} \\ \\ 
f_{d} 
 \end{Bmatrix}+\begin{bmatrix}
I & \phantom{-}0 \\ \\ 
\,0 & \phantom{-}\omega^{2}M_{dd} \\ \\ 
 \end{bmatrix}\begin{Bmatrix}
0 \\ \\ 
\Delta_{d} 
 \end{Bmatrix} +P\begin{bmatrix}
0 & \phantom{-}0 \\ \\ 
G_{db} & \phantom{-}G_{dd} \\ \\ 
 \end{bmatrix}\begin{Bmatrix}
\Delta_{b} \\ \\ 
\Delta_{d} 
 \end{Bmatrix} \tag{43}
\end{align*}

\noindent Here the vector $\Delta_{b}$ contains the boundary related non-zero slope, curvature and triple displacement derivative dofs. Similarly, the vector $\Delta_{d}$ includes all the non-zero displacement dofs of the beam element. For bending analysis $M_{dd}=P=0$, for free vibration analysis $f_{b}=f_{d}=P=0$ and  $f_{b}=f_{d}=M_{dd}=0$ for stability analysis. The solution of the Equation (\ref{eq:Boundary_disp_Beam}) after applying the appropriate boundary conditions gives the unknown displacements, frequencies and buckling load.
\section{Numerical Results and Discussion}

The performance of the proposed two quadrature beam element is assessed through static, free vibration and stability analysis of second strain gradient Euler-Bernoulli beam. In the present analysis single DQ element is used to model the prismatic beams under different boundary and loading conditions. The element based on Lagrange interpolation is designated as SQE8-L and the element based on Hermite interpolation by SQE8-H. The results reported herein are obtained using single quadrature element for $g_{1}=0.015, g_{2}=0.01$ and compared with the analytical values computed in the Appendix-I. The numerical data used for the analysis of beams is as follows: Length $L=1$, Young's modulus $E=3 \times 10^{6}$, Poission's ratio $\nu=0.3$, density $\rho=1$ and transverse load $q=1$.\\
\subsection{Static analysis of second strain gradient beam using quadrature elements}

A simply support beam subjected to uniformly distributed load is considered in the present analysis. The deflection reported here is nondimensional as, : $\bar{w}=100EIw/qL^{4}$. In Table \ref{Defl_ss_udl}, convergence of nondimensional deflection and slope obtained using SQE8-L and SQE8-H are compared with exact solutions obtained in Appendix-I. The deflection is evaluated at center of the beam $x=0$ and slope are computed at the left support $x=-L/2$. Excellent convergence is exhibited by both elements for deflection and slope using 11 grid points. In Table \ref{Defl_ss_beam_udl_along_length}, the nondimensional deflection along the length of a simply suported beam obtained using 11 noded SQE8-L and SQE8-H elements are presented. The results show perfect match with the analytical values along the full length of the beam.

\setlength{\extrarowheight}{.5em}
\begin{table}[H]
    \centering
    \caption{Comparison of deflection and slope for a simply supported beam under a udl.}

\begin{tabular}{|c@{\hskip 0.15in}|c@{\hskip 0.15in}c@{\hskip 0.15in}|c@{\hskip 0.15in}c@{\hskip 0.15in}|}
\hline
\multirow{3}{*}{N} & \multicolumn{2}{c|}{${w}_{\,(x=0)}$} & %
 \multicolumn{2}{c|}{$w^{'}_{\,(x=-L/2)}$ } \\
\cline{2-5}
 & SQE8-L & SQE8-H & SQE8-L & SQE8-H  \\
\cline{1-5}

$7$   &  1.2992  & 1.6391 & 0.1667   & 0.1724  \\ 

$9$   &  1.2993   &  1.2981 & 0.1667  & 0.1660 \\ 

$11$    &  1.2993 & 1.2992 & 0.1666   & 0.1659 \\

 $13$    &   1.2993  & 1.2992 &  0.1666    &  0.1659  \\

$15$     &   1.2993   & 1.2992 & 0.1666   & 0.1659 \\ 
$17$     &   1.2993  &  1.2992 & 0.1665  & 0.1660\\ 
$19$     &  1.2993  & 1.2992 &  0.1665   &  0.1660\\ 
$21$     &  1.2993   &  1.2992 & 0.1665  &  0.1660 \\  
 $\text{Exact}$   & 1.2992      &  & 0.1660 &  \\ \hline 
 
\end{tabular}

    \label{Defl_ss_udl}
\end{table}

\setlength{\extrarowheight}{.5em}
\begin{table}[H]
    \centering
    \caption{Comparison of deflection along the length for a simply supported  beam under a udl.}

\begin{tabular}{|c@{\hskip 0.15in}|c@{\hskip 0.15in}|c@{\hskip 0.175in}|c@{\hskip 0.175in}|}
\hline
$x=2\xi/L$ & SQE8-L & SQE8-H & Exact\\ \hline
 -1.0000   &  0.0000  &  0.0000  & 0.0000  \\
 -0.9340& 0.1369 &  0.1367 &  0.1367   \\ 
-0.7845	&   0.4381 & 0.4379 &  0.4379\\
-0.5652&   0.8276&  0.8272  & 0.8274  \\  
  -0.2957 &  1.1649 & 1.1646 & 1.1647  \\
 0.0000	 &   1.2993 &  1.2992 & 1.2992 \\
 0.2957	 &   1.1649 &  1.1646 & 1.1647  \\
0.5652&  0.8276&  0.8272  & 0.8274     \\  
0.7845	&   0.4381 & 0.4379 &  0.4379 \\
0.9340& 0.1369 &  0.1367 &  0.1367   \\
1.0000   &  0.0000  &  0.0000  & 0.0000   \\
\hline 
 
\end{tabular}

    \label{Defl_ss_beam_udl_along_length}
\end{table}

From the above observations it is concluded that, both SQE8-L and SQE8-H elements can be efficiently applied to study the static behaviour of second strain gradient Euler-Bernoulli beam with less number of nodes.

\subsubsection{Free vibration analysis of second gradient beam using quadrature elements} 
 
 All the frequencies reported for beams are nondimensional as $\bar\omega=\omega{L^{2}\sqrt{\rho{A}/EI}}$. The analytical solutions for second strain gradient elastic Euler-Bernoulli beam with different boundary conditions are obtained by following the approach given in \cite{Kitar} and the associated frequency equations are presented in Appendix-I.  In this study, the rotary inertia related to slope, curvature and triple displacement derivative dofs is neglected. In Table \ref{Freq_ss_beam}, convergence of first six frequencies for a simply supported beam obtained using the SQE8-L and SQE8-H elements for $g_{1}=0.015, g_{2}=0.01$ are presented and compared with analytical values. The rate of convergence is seen faster for the both the elements for all the frequencies. Similar trend is noticed in the the Table \ref{Freq_free_free_beam}, for a free-free beam. \\
Hence, a single SQE8-L or SQE8-H element with less number of nodes can produce accurate solutions even for higher frequencies and can be efficiently applied for free vibration analysis of second strain gradient beams.
 
\setlength{\extrarowheight}{.5em}
\begin{table}[H]
    \centering
    \caption{Comparison of first six frequencies for a simply supported beam.}
      
 \begin{tabular}{c@{\hskip 0.1in}c@{\hskip 0.1in}c@{\hskip 0.1in}c@{\hskip 0.1in}c@{\hskip 0.1in}c@{\hskip 0.1in}c@{\hskip 0.1in}c@{\hskip 0.1in}}
     \\ \hline
N & Model & $\bar{\omega}_{1}$    & $\bar{\omega}_{2}$  & $\bar{\omega}_{3}$& $\bar{\omega}_{4}$ &$\bar{\omega}_{5}$ &$\bar{\omega}_{6}$
\\ \hline 
  
7&SQE8-L&  9.8894 & 39.6754  &  86.7154  & 205.0036  & 230.0011 &--- \\ 
&SQE8-H&  8.8946 & 33.2850  & 55.5569  & 135.7379  & 263.6469 &--- \\ 

9&SQE8-L&9.8804 & 39.6511 & 89.6556  & 162.8854 &  230.4984 &  607.4897  \\ 
&SQE8-H&9.8814 & 39.6565 & 73.9288  & 135.9277 &  223.6142 &  505.4761  \\ 

11& SQE8-L&9.8803 	 & 39.6501 &  89.6955 & 160.7182  & 251.6413 & 394.3007   \\
&SQE8-H &9.8814 	 & 39.6658 &  89.2624 & 159.4814  & 219.0450 & 353.6155 \\ 

13&SQE8-L&9.8803 	 & 39.6494 &  89.6916 & 160.6484  & 253.4009 & 371.6141   \\
&SQE8-H&9.8812 	 & 39.6648 &  89.7535 & 160.7850  & 248.9904 & 365.4424 \\ 

15&SQE8-L&9.8803 	 & 39.6490 &  89.6894 & 160.6407  & 253.3872 & 369.3019   \\
&SQE8-H &9.8811 	 & 39.6624 &  89.7571 & 160.8537  & 253.6615 & 369.5068 \\ 

17& SQE8-L&9.8802 	 & 39.6490 &  89.6888 & 160.6399  & 253.3860 & 369.0764  \\
&SQE8-H &9.8810 	 & 39.6611 &  89.7504 & 160.8307  & 253.8495 & 370.0311 \\ 

19&SQE8-L&9.8802 	 & 39.6491 &  89.6892 & 160.6422  & 253.3883 & 369.0978  \\
&SQE8-H & 9.8810 	 & 39.6604 &  89.7472 & 160.8209  & 253.8282 & 369.9803 \\ 

21& SQE8-L&9.8802 	 & 39.6494 &  89.6901 & 160.6460  & 253.3951 & 369.1136  \\
&SQE8-H& 9.8810 	 & 39.6602 &  89.7459 & 160.8169  & 253.8183 & 369.9609 \\ 
& Analytical & 9.8810 	 & 39.6600 &  89.7454 & 160.8149  & 253.8140 & 369.9512  	 \\ \hline 

        \end{tabular}
    \label{Freq_ss_beam}
\end{table}

\setlength{\extrarowheight}{.5em}
\begin{table}[H]
    \centering
    \caption{Comparison of first six elastic frequencies for a free-free beam.}
      
 \begin{tabular}{c@{\hskip 0.1in}c@{\hskip 0.1in}c@{\hskip 0.1in}c@{\hskip 0.1in}c@{\hskip 0.1in}c@{\hskip 0.1in}c@{\hskip 0.1in}c@{\hskip 0.1in}}
     \\ \hline
N & Model & $\bar{\omega}_{1}$    & $\bar{\omega}_{2}$  & $\bar{\omega}_{3}$& $\bar{\omega}_{4}$ &$\bar{\omega}_{5}$ &$\bar{\omega}_{6}$
\\ \hline 
  
7&SQE8-L &  22.4033 	 & 62.3142 &  111.4149 & 340.7851 & 353.7644 &--- \\ 
&SQE8-H& 17.5096	 & 43.8559 &  88.2238 & 251.2650 & 356.4709 &--- \\ 

9&SQE8-L & 22.4040 	 & 61.9927 &  121.8060 & 212.6738 & 268.6578 & 988.6651  \\ 
&SQE8-H & 22.2382 	 & 60.5383 &  93.3097 & 169.9119 & 287.7418 & 880.3811 \\ 

11& SQE8-L & 22.4040 	 & 61.9898 &  122.2289 & 204.1210 & 301.0061 & 497.0841 \\
&SQE8-H & 22.4040 	 & 61.9839 &  118.1267 & 195.3483 & 268.7911 & 451.4562 \\ 

13&SQE8-L & 22.4040 	 & 61.9899 &  122.2347 & 203.6668 & 307.0701 & 434.3626 \\
&SQE8-H & 22.4040 	 & 61.9899 &  122.1750 & 203.4007 & 294.1555 & 424.9945 \\ 

15&SQE8-L&22.4040 	 & 61.9899 &  122.2343 & 203.6670 & 307.2807 & 434.8845 \\
&SQE8-H &22.4040 	 & 61.9899 &  122.2343 & 203.6670 & 307.2807 & 434.8845 \\ 

17& SQE8-L& 22.4040 	 & 61.9899 &  122.2343 & 203.6644 & 307.2691 & 434.3392 \\
&SQE8-H & 22.4040 	 & 61.9899 &  122.2350 & 203.6668 & 307.2620 & 434.2469 \\ 

19&SQE8-L & 22.4040 	 & 61.9899 &  122.2344 & 203.6645 & 307.2707 & 434.2732 \\
&SQE8-H & 22.4040 	 & 61.9899 &  122.2350 & 203.6667 & 307.2777 & 434.2899 \\ 

21& SQE8-L& 22.4040 	 & 61.9899 &  122.2345 & 203.6647 & 307.2719 & 434.2764 \\
&SQE8-H& 22.4040 	 & 61.9900 &  122.2350 & 203.6667 & 307.2779 & 434.2905 \\ 

& Analytical & 22.4040 	 & 61.9900 &  122.2350 & 203.6676& 307.2790 & 434.2906	 \\ \hline 

        \end{tabular}
    \label{Freq_free_free_beam}
\end{table}

\subsection{Stability analysis of gradient elastic beams using quadrature elements}

In this section, the applicability of the  proposed elements for stability analysis of prismatic second gradient Euler-Bernoulli beam is verified. The convergence of the critical buckling load for a simply supported beam obtained using SQE8-L and SQE8-H is shown in Table \ref{Pcr_ss}, and compared with the analytical values obtained in the Appendix-I for $g_{1}=0.015, g_{2}=0.01$. Good convergence trend is exhibited by both the elements. Hence, a single SQE8-L or SQE8-H element with few nodes can be efficiently used to  study the buckling behaviour of a second strain gradient elastic beam. 
\setlength{\extrarowheight}{.5em}
\begin{table}
    \centering
    \caption{Comparison of normalized buckling load for a simply supported beam.}

\begin{tabular}{|c@{\hskip 0.2in}|c@{\hskip 0.2in}|c@{\hskip 0.2in}|} 
\hline
$ \text{N}$ & SQE8-L &  SQE8-H	 \\
 \hline 
 $5$	 &9.8842       &   15.1639  \\  
  $6$	 & 9.8915   &   8.9866  \\  
 $7$ &    9.8913 &  9.6074  \\ 
 $8$ &   9.8912  &  9.9038   \\  
$9$	&   9.8912 &   9.9048 \\
$10$	& 9.8911  &  9.8935 \\
 $11$	 &   9.8912       & 9.8931  \\  
  $12$	 &   9.8910       & 9.8931  \\  
   $13$	 &   9.8910       & 9.8929  \\  
    $14$	 &   9.8910       & 9.8927  \\  
     $15$	 &   9.8909       & 9.8926  \\  
Analytical  &   9.8926    &  \\  
\hline

\end{tabular}
    \label{Pcr_ss}
\end{table}

\section{Conclusion}

Two novel versions of weak form quadrature elements are proposed to solve a eighth order differential equation associated with the second strain gradient Euler-Bernoulli beam theory. The two elements are based on Lagrange and Hermite interpolations, respectively. A novel way was introduced to account for the multi-degrees of freedom related to second strain gradient theory. The performance of the proposed elements was demonstrated through numerical examples on bending, free vibration and stability analysis of prismatic gradient beams. Based on the findings it was concluded that, both elements exhibit excellent performance with fewer number of nodes.

\section*{APPENDIX}

\section*{Analytical solutions for second strain gradient Euler-Bernoulli beam}

In this section we obtain the analytical solutions for bending, free vibration and stability analysis of second strain gradient Euler-Bernoulli beam. 

\subsection*{Bending analysis}
Let us consider a beam of length \textit{L} subjected to a uniformly distributed load \textit{q}. To obtain the static deflections of the second gradient elastic Euler-Bernoulli beam which is governed by Equation (\ref{EOM_beam}), we assume a solution of the form

 \begin{align*}  \label{disp_exp_exact}       
w(x)= c_{1}+c_{2}x+c_{3}x^{2}+c_{4}x^{3}+c_{5}e^{n_{1}x}+
c_{6}e^{n_{2}x}+c_{7}e^{m_{1}x}+c_{8}e^{m_{2}x}-\frac{qx^{4}}{24EI} \tag{A1}
\end{align*} 

where 
 \begin{align*}         
n_{1}=\sqrt{\frac{g_{1}^{2}+\sqrt{g_{1}^{4}-4g_{2}^{4}}}{2g_{2}^{4}}}, \quad
n_{2}=-\sqrt{\frac{g_{1}^{2}+\sqrt{g_{1}^{4}-4g_{2}^{4}}}{2g_{2}^{4}}}, \quad \\ m_{1}=\sqrt{\frac{g_{1}^{2}-\sqrt{g_{1}^{4}-4g_{2}^{4}}}{2g_{2}^{4}}}, \quad
m_{2}=-\sqrt{\frac{g_{1}^{2}-\sqrt{g_{1}^{4}-4g_{2}^{4}}}{2g_{2}^{4}}}, \quad
\end{align*}

\noindent The constants $c_{1}-c_{8}$ are determined with the aid of boundary conditions listed in Equation (\ref{eq:BC_Cl_Beam}) and (\ref{eq:BC_NCl_Beam}). After applying the boundary conditions the system of equations are expressed as:

\begin{align*}        
[K]\{\delta\}=\{f\} \tag{A2}
\end{align*}

here $K$ is the coefficient matrix, $f$ is the vector corresponding to the load and $\{\delta\}=\{c_{1},c_{2},c_{3},c_{4},c_{5},c_{6},c_{7},c_{8}\}$ is the unknown constant vector to be determined. Once the unknown constants are obtained then the displacement solution is computed from the Equation (A1). The slope at any point along the length of the beam can be obtained by performing the first derivatives of the displacement. To have real and positive roots we assumed $g_{1}/g_{2} > \sqrt{2}$ in the present analysis. The simultaneous equations to determine the unknown coefficients for a simply supported beam is given as: \\

\noindent (a) Simply supported beam :
$$[K]=
\begin{bmatrix}
1 & 0 & 0 & 0 & 1 & 1 & 1 & 1\\
1 & L & L^2 & L^3 & e^{m_{1}L} & e^{m_{2}L} & e{n_{1}L} & e^{n_{2}L}\\
0 & 0 & 2 & 0 & a_{11} & a_{12}  &
a_{13} & a_{14}\\
    0 & 0 & 2 & 6L & b_{11} & b_{12} &
b_{13} & b_{14}\\  
     0 & 0 & 2 & 0 & m_{1}^{2} & m_{2}^{2} & n_{1}^{2}  & n_{2}^{2} \\
     0 & 0 & 2 & 6L & m_{1}^{2}e^{m_{1}L} & m_{2}^{2}e^{m_{2}L} & n_{1}^{2}e^{n_{1}L} & n_{2}^{2}e^{n_{2}L}\\
     0 & 0 & 0 & 6 & m_{1}^{3} & m_{2}^{3} & n_{1}^{3} & n_{2}^{3}\\
     0 & 0 & 0 & 6 & m_{1}^{3}e^{m_{1}L} & m_{2}^{3}e^{m_{2}L} & n_{1}^{3}e^{n_{1}L} & n_{2}^{3}e^{n_{2}L}\\
     \end{bmatrix}, \,\,\, \{f\}=\begin{Bmatrix}
 0 \\ 
-qL^{4}/24EI \\ 
g_{1}^{2}q/EI\\ 
g_{1}^{2}q/EI-qL^{2}/2EI \\  
 0 \\ 
-qL^{2}/2EI\\ 
0\\
-qL/EI\\  
\end{Bmatrix}
 $$ \\ \
 
\noindent where, \\
\noindent $ a_{11}= m_{1}^{2}-g_{1}^{2}m_{1}^{4}+g_{2}^{4}m_{1}^{6},\,\,\, \quad a_{12}= m_{2}^{2}-g_{1}^{2}m_{2}^{4}+g_{2}^{4}m_{2}^{6},\\ a_{13}=n_{1}^{2}-g_{1}^{2}n_{1}^{4}+g_{2}^{4}n_{1}^{6}\,\,, \,\,\,\,\,a_{14}= n_{2}^{2}-g_{1}^{2}n_{2}^{4}+g_{2}^{4}n_{2}^{6}\\ \\
b_{11}= (m_{1}^{2}-g_{1}^{2}m_{1}^{4}+g_{2}^{4}m_{1}^{6})e^{m_{1}L}, \,\,\,b_{12}= (m_{2}^{2}-g_{1}^{2}m_{2}^{4}+g_{2}^{4}m_{2}^{6})e^{m_{2}L},
\\ b_{13}= (n_{1}^{2}-g_{1}^{2}n_{1}^{4}+g_{2}^{4}n_{1}^{6})e^{n_{1}L},\,\,\,\,\,\,\,\,b_{14}= (n_{2}^{2}-g_{1}^{2}n_{2}^{4}+g_{2}^{4}n_{2}^{6})e^{n_{2}L}
$ \\

\subsection*{Free vibration analysis}

To obtain the natural frequencies for a second gradient elastic Euler-Bernoulli beam which is governed by Equation (\ref{EOM_beam}), we assume a solution of the form

\begin{align*}         
w(x,t)=\bar{w}(x){e}^{i\omega{t}}  \tag{B1}
\end{align*}

\noindent substituting the above solution in the governing equation (\ref{EOM_beam}), we get

\begin{align*}         
\bar{w}^{iv}-g_{1}^{2}\bar{w}^{vi}+g_{2}^{4}\bar{w}^{viii}-\frac{\omega^{2}}{\beta^{2}}\bar{w}=0  \tag{B2}
\end{align*}

\noindent here, $\beta^{2}=EI/m$, and the above equation has the solution of type

\begin{align*}         
\bar{w}(x)=\sum_{j=1}^{8}c_{i}{e}^{k_{i}x}  \tag{B3}
\end{align*}

\noindent where, $c_{i}$ are the constants of integration which are determined through boundary conditions and the $k_{i}$ are the roots of the characteristic equation 

\begin{align*}         
{k}^{iv}-g_{1}^{2}{k}^{vi}+g_{2}^{4}{k}^{viii}-\frac{\omega^{2}}{\beta^{2}}=0 \tag{B4}
\end{align*}

After applying the boundary conditions listed in Equations (\ref{eq:BC_Cl_Beam}) and (\ref{eq:BC_NCl_Beam}) we get,
\begin{align*}         
[F(\omega)]\{C\}=\{0\} \tag{B5}
\end{align*}

For non-trivial solution, following condition should be satisfied

\begin{align*}         
det[F(\omega)]=0 \tag{B6}
\end{align*}

The above frequency equation renders all the natural frequencies for a second strain gradient Euler-Bernoulli beam. The following are the frequency equations for simply supported and free-free boundary conditions.\\

\noindent (a) Simply supported beam :
\begin{align*}\label{exact_freq_ss_EQ}
[F(\omega)]=\begin{bmatrix}
1 & 1 & 1 & 1 & 1 & 1& 1 & 1\\
{e}^{(k_{1}L)} & {e}^{(k_{2}L)} & {e}^{(k_{3}L)} & {e}^{(k_{4}L)} & {e}^{(k_{5}L)} & {e}^{(k_{6}L)}& {e}^{(k_{7}L)} & {e}^{(k_{8}L)} \\
{k_{1}}^2 & {k_{2}}^2 & {k_{3}}^2 & {k_{4}}^2 & {k_{5}}^2 & {k_{6}}^2& {k_{7}}^2 & {k_{8}}^2\\ 
t_{1} &t_{2}&t_{3}&t_{4}&t_{5}&t_{6}&t_{7}&t_{8}\\ 
t_{1}{e}^{(k_{1}L)} &t_{2}{e}^{(k_{2}L)}&t_{3}{e}^{(k_{3}L)}&t_{4}{e}^{(k_{4}L)}&t_{5}{e}^{(k_{5}L)}&t_{6}{e}^{(k_{6}L)}&t_{7}{e}^{(k_{7}L)}&t_{8}{e}^{(k_{8}L)}\\
k_{1}^{3} & k_{2}^{3} & k_{3}^{3} & k_{4}^{3} & k_{5}^{3} & k_{6}^{3}& k_{7}^{3} & k_{8}^{3}  \\
k_{1}^{3}{e}^{(k_{1}L)} & k_{2}^{3}{e}^{(k_{2}L)} & k_{3}^{3}{e}^{(k_{3}L)} & k_{4}^{3}{e}^{(k_{4}L)} & k_{5}^{3}{e}^{(k_{5}L)} & k_{6}^{3}{e}^{(k_{6}L)} &  k_{7}^{3}{e}^{(k_{7}L)} & k_{8}^{3}{e}^{(k_{8}L)}\\
\end{bmatrix} 
\end{align*}
\noindent (e) Free-free beam :
\begin{align*}
[F(\omega)]=
\begin{bmatrix}
p_{1} &p_{2}&p_{3}&p_{4}&p_{5}&p_{6}&p_{7}&p_{8}\\ 
r_{1} &r_{2}&r_{3}&r_{4}&r_{5}&r_{6}&r_{7}&r_{8}\\ 
q_{1} &q_{2}&q_{3}&p_{4}&q_{5}&q_{6}&q_{7}&q_{8}\\ 
k_{1}^{4} & k_{2}^{4} & k_{3}^{4} & k_{4}^{4} & k_{5}^{4} & k_{6}^{4}&  k_{7}^{4} & k_{8}^{4}\\
p_{1}{e}^{(k_{1}L)} &p_{2}{e}^{(k_{2}L)}&p_{3}{e}^{(k_{3}L)}&p_{4}{e}^{(k_{4}L)}&p_{5}{e}^{(k_{5}L)}&p_{6}{e}^{(k_{6}L)}&p_{7}{e}^{(k_{7}L)}&p_{8}{e}^{(k_{8}L)}\\ 
r_{1}{e}^{(k_{1}L)} &r_{2}{e}^{(k_{2}L)}&r_{3}{e}^{(k_{3}L)}&r_{4}{e}^{(k_{4}L)}&r_{5}{e}^{(k_{5}L)}&r_{6}{e}^{(k_{6}L)}&r_{7}{e}^{(k_{7}L)}&r_{8}{e}^{(k_{8}L)}\\ 
q_{1}{e}^{(k_{1}L)} &q_{2}{e}^{(k_{2}L)}&q_{3}{e}^{(k_{3}L)}&p_{4}{e}^{(k_{4}L)}&q_{5}{e}^{(k_{5}L)}&q_{6}{e}^{(k_{6}L)}&s_{7}{e}^{(k_{7}L)}&q_{8}{e}^{(k_{8}L)}\\ 
k_{1}^{4}{e}^{(k_{1}L)} & k_{2}^{4}{e}^{(k_{2}L)} & k_{3}^{4}{e}^{(k_{3}L)} & k_{4}^{4}{e}^{(k_{4}L)} & k_{5}^{4}{e}^{(k_{5}L)} & k_{6}^{4}{e}^{(k_{6}L)} &  k_{7}^{4}{e}^{(k_{7}L)} & k_{8}^{4}{e}^{(k_{8}L)}\\
\end{bmatrix} 
\end{align*}
\noindent Where,  \\

\noindent $t_{1}=(-g_{1}^{2}k_{1}^{4}+g_{2}^{4}k_{1}^{6}),\quad  t_{2}=(-g_{1}^{2}k_{2}^{4}+g_{2}^{4}k_{2}^{6}) ,\quad t_{3}=(-g_{1}^{2}k_{3}^{4}+g_{2}^{4}k_{3}^{6})$ \\ 
$t_{4}=(-g_{1}^{2}k_{4}^{4}+g_{2}^{4}k_{4}^{6}) ,\quad t_{5}=(-g_{1}^{2}k_{5}^{4}+g_{2}^{4}k_{5}^{6}) \quad t_{6}=(-g_{1}^{2}k_{6}^{4}+g_{2}^{4}k_{6}^{6})$ \\
$t_{7}=(-g_{1}^{2}k_{7}^{4}+g_{2}^{4}k_{7}^{6}) ,\quad t_{8}=(-g_{1}^{2}k_{8}^{4}+g_{2}^{4}k_{8}^{6})$ \\

\noindent $p_{1}=(k_{1}^{3}-g_{1}^{2}{k_{1}}^{5}+g_{2}^{4}{k_{1}}^{7}) ,\quad  p_{2}=(k_{2}^{3}-g_{1}^{2}{k_{2}}^{5}+g_{2}^{4}{k_{2}}^{7})\quad \\ p_{3}=(k_{3}^{3}-g_{1}^{2}{k_{3}}^{5}+g_{2}^{4}{k_{3}}^{7})$ , \,\, 
$p_{4}=(k_{4}^{3}-g_{1}^{2}{k_{4}}^{5}+g_{2}^{4}{k_{4}}^{7}) \\ \quad p_{5}=(k_{5}^{3}-g_{1}^{2}{k_{5}}^{5}+g_{2}^{4}{k_{5}}^{7}), \quad p_{6}=(k_{6}^{3}-g_{1}^{2}{k_{6}}^{5}+g_{2}^{4}{k_{6}}^{7}) \\ p_{7}=(k_{7}^{3}-g_{1}^{2}{k_{7}}^{5}+g_{2}^{4}{k_{7}}^{7}), \quad p_{8}=(k_{8}^{3}-g_{1}^{2}{k_{8}}^{5}+g_{2}^{4}{k_{8}}^{7})$ \\

\noindent $r_{1}=(k_{1}^{2}-g_{1}^{2}{k_{1}}^{4}+g_{2}^{4}{k_{1}}^{6}) ,\quad  r_{2}=(k_{2}^{2}-g_{1}^{2}{k_{2}}^{4}+g_{2}^{4}{k_{2}}^{6}) \quad \\ r_{3}=(k_{3}^{2}-g_{1}^{2}{k_{3}}^{4}+g_{2}^{4}{k_{3}}^{6})$ , \,\, 
$r_{4}=(k_{4}^{2}-g_{1}^{2}{k_{4}}^{4}+g_{2}^{4}{k_{4}}^{6})  \\ \quad r_{5}=(k_{5}^{2}-g_{1}^{2}{k_{5}}^{4}+g_{2}^{4}{k_{5}}^{6}), \quad r_{6}=(k_{6}^{2}-g_{1}^{2}{k_{6}}^{4}+g_{2}^{4}{k_{6}}^{6}) \\ r_{7}=(k_{7}^{2}-g_{1}^{2}{k_{7}}^{4}+g_{2}^{4}{k_{7}}^{6}), \quad r_{8}=(k_{8}^{2}-g_{1}^{2}{k_{8}}^{4}+g_{2}^{4}{k_{8}}^{6})$ \\

\noindent $q_{1}=(g_{1}^{2}{k_{1}}^{3}-g_{2}^{4}{k_{1}}^{5}) ,\quad  q_{2}=(g_{1}^{2}{k_{2}}^{3}-g_{2}^{4}{k_{2}}^{5}) \quad \\ q_{3}=(g_{1}^{2}{k_{3}}^{3}-g_{2}^{4}{k_{3}}^{5})$ , \,\, 
$q_{4}=(g_{1}^{2}{k_{4}}^{3}-g_{2}^{4}{k_{4}}^{5})  \\ \quad q_{5}=(g_{1}^{2}{k_{5}}^{3}-g_{2}^{4}{k_{5}}^{5}), \quad q_{6}=(g_{1}^{2}{k_{6}}^{3}-g_{2}^{4}{k_{6}}^{5}) \\ q_{7}=(g_{1}^{2}{k_{7}}^{3}-g_{2}^{4}{k_{7}}^{5}), \quad q_{8}=(g_{1}^{2}{k_{8}}^{3}-g_{2}^{4}{k_{8}}^{5})$ \\

\subsection*{Stability analysis}

To obtain the buckling load for a second strain gradient Euler-Bernoulli beam which is governed by Equation (\ref{EOM_beam}), we assume a solution of the form

\begin{align*}         
w(x)=c_{1}+c_{2}x+c_{3}e^{m_{1}x}+c_{4}e^{m_{2}x}+c_{5}e^{m_{3}x}+
c_{6}e^{n_{1}x}+c_{7}e^{n_{2}x}+c_{8}e^{n_{3}x} \\ \tag{C1}
\end{align*}

\noindent where, $c_{i}$ are the constants of integration which are determined using the boundary conditions and $m_{1,2,3}$ and $n_{1,2,3}$ are the roots of the following characteristic equation:
\begin{align*}         
g_{2}^{4}s^{6}-g_{1}^{2}s^{4}+s^{2}+\frac{P}{EI}=0   \tag{C2}
\end{align*}

After applying the boundary conditions listed in Equations (\ref{eq:BC_Cl_Beam}) and (\ref{eq:BC_NCl_Beam}) we get,
\begin{align*}         
[\bar{G}(P)]\{C\}=\{0\}   \tag{C3}
\end{align*}

For non-trivial solution,
\begin{align*}         
det[\bar{G}(P)]=0    \tag{C4}
\end{align*}

The above Eigenvalue problem yields the buckling load for a second strain gradient Euler-Bernoulli beam. The system equations for a simply supported beam is given as:\\
 
\noindent (a) Simply supported beam :
$$[F(\omega)]=
\begin{bmatrix}
1 & 0 & 1 & 1 & 1 & 1& 1 & 1\\
1 & L & {e}^{(m_{1}L)} & {e}^{(m_{2}L)} & {e}^{(m_{3}L)} & {e}^{(n_{1}L)}& {e}^{(n_{2}L)} & {e}^{(n_{3}L)} \\
0 & 0 & {m_{1}}^2 & {m_{2}}^2 & {m_{3}}^2 & {n_{1}}^2& {n_{2}}^2 & {n_{3}}^2\\ 
0 & 0 & t_{3}&t_{4}&t_{5}&t_{6}&t_{7}&t_{8}\\ 
0 & 0 & t_{3}{e}^{(m_{1}L)}&t_{4}{e}^{(m_{2}L)}&t_{5}{e}^{(m_{3}L)}&t_{6}{e}^{(n_{1}L)}&t_{7}{e}^{(n_{2}L)}&t_{8}{e}^{(n_{3}L)}\\
0 & 0 & m_{1}^{3} & m_{2}^{3} & m_{3}^{3} & n_{1}^{3}& n_{2}^{3} & n_{3}^{3}  \\
0 & 0 & m_{1}^{3}{e}^{(m_{1}L)} & m_{2}^{3}{e}^{(m_{2}L)} & n_{1}^{3}{e}^{(n_{1}L)} & n_{2}^{3}{e}^{(n_{2}L)} &  n_{3}^{3}{e}^{(n_{3}L)} & n_{4}^{3}{e}^{(n_{4}L)}\\
\end{bmatrix}  
$$ \\
\end{document}